\documentclass[12pt]{article}
\usepackage{epsfig}
\usepackage{subfigure}
\usepackage{graphicx}

\begin{document}

\begin{titlepage}
\begin{center}
\hfill    CERN-TH-2016-039\\

\vskip 1cm

{\large \bf {What if the Masses of the First Two Quark Families are not Generated by 
the Standard Higgs?}}

\vskip 1cm

F. J. Botella  $^{a,c}$ \footnote{fbotella@uv.es}, 
G. C. Branco  $^{b, c}$ \footnote{gbranco@tecnico.ulisboa.pt}, 
M. N. Rebelo $^{b, c}$ \footnote{rebelo@tecnico.ulisboa.pt},
and J. I. Silva-Marcos $^b$\footnote{juca@cftp.ist.utl.pt},

\vspace{1.0cm}

{\it $^a$ Departament de F\' \i sica Te\`orica and IFIC,
Universitat de Val\`encia-CSIC, E-46100, Burjassot, Spain.} \\
{\it $^b$ Departamento de F\'\i sica and Centro de F\' \i sica Te\' orica
de Part\' \i culas (CFTP),
Instituto Superior T\' ecnico (IST), U. de Lisboa (UL), Av. Rovisco Pais, P-1049-001 
Lisboa, Portugal. \\
\it$^c$ Theory Department, CERN, CH 1211 Geneva 23, Switzerland}

\end{center}

\vskip 3cm

\begin{abstract}
We point out that, in the context of the SM, $|V^2_{13}|  + | V^2_{23}|$ is expected
to be large, of order one. The fact that $|V^2_{13}|  +  |V^2_{23}| \approx  1.6 \times 10^{-3}$ 
motivates the introduction of a symmetry S which leads to $V_{CKM} ={1\>\!\!\!\mathrm{I}} $,
with only the third generation of quarks acquiring mass. We consider two scenarios 
for generating the mass of the first two quark generations and full quark mixing. One consists 
of the introduction  of a second Higgs doublet which is neutral under S. The second scenario 
consists of assuming 
New Physics at a high energy scale , contributing to the masses of light quark generations, in an 
effective field theory approach. This last scenario leads to couplings of the Higgs particle to 
$s\overline s$  and $c \overline c$ which are significantly enhanced with respect 
to  those of the SM. In both schemes, one has scalar-mediated flavour- changing neutral 
currents which are naturally suppressed. Flavour violating top decays are predicted
in the second scenario at the level $ \mbox{Br} (t \rightarrow h c )
\geq 5\times 10^{-5}$.

\end{abstract}

\end{titlepage}

\newpage

The recent discovery of the Higgs particle at LHC, rendered even more urgent
to understand the mechanism responsible for the generation of fermion masses
and mixing. In the framework of the Standard Model (SM), fermion masses
arise exclusively through Yukawa interactions and the Brout-Englert-Higgs
mechanism is responsible for both gauge symmetry breaking and the generation
of fermion masses. Some of the outstanding questions one may ask, include :

i) Two of the salient flavour features in the quark sector are the strong
hierarchy of quark masses and the fact that the $V_{CKM}$ matrix is close to
the identity. In the framework of the SM, can one conclude that these two
features are related in some way? How can one understand small quark mixing
in the SM ?

ii) In the SM, all fermion masses are generated through the vacuum
expectation value (vev) of the Standard Higgs. Alternatively, one may
consider a scenario where the Standard Higgs only gives mass to the third
generation, while the masses of the two first generations originate from
another source. A crucial question is : how can this alternative scenario be
tested at LHC and future accelerators?

In this paper, we address the above two questions. With respect to (i), we
show that actually in the SM the ``natural" value of $(|V^2_{13}| + |
V^2_{23}|)$ is large, of order one. In order to address this question, we
study in detail quark mixing in the extreme chiral (EC) limit, where only
the third generation of quarks acquires mass, while $m_d$, $m_s$, $m_u$, $%
m_c $ remain massless. We do the analysis in the context of the Standard
Model (SM) and some of its extensions. We will show that in the SM in the EC
limit the generic situation is having non-trivial mixing parametrised by an
angle with a free value, not fixed in the SM context. Without loss of
generality, one can identify this angle with the $V_{23}$ entry. Therefore,
the fact that experimentally $|V_{23}| = 4.09 \times 10^{-2}$, is
entirely unnatural within the framework of the SM. In fact, the smallness
of $|V_{23}|$ may be interpreted as a hint from experiment, indicating that
one should find a symmetry or a principle which may account for the
smallness of $|V_{23}|$.

In our analysis, we start with the most general rank one matrices $M_u$, $
M_d $, taking into account that in the SM the flavour structure of the
Yukawa couplings generating the up and down quark mass matrices are entirely
independent. The appearance of a non-trivial mixing even in the EC limit
case, corresponds to a misalignment of the two mass matrices $M_u$, $M_d$,
in flavour space. We define a dimensionless weak basis (WB) invariant
denoted A which provides a measure of this misalignment. In the EC limit,
this invariant A varies from 0 to 1 , with 0 corresponding to exact
alignment and 1 to total misalignment.

With respect to question (ii) we consider the possibility that in leading
order the SM Higgs only gives mass to the third generation. This is achieved
in a natural way through the introduction of a discrete symmetry S which
leads to quark mass matrices of rank one, aligned in flavour space. We then
conjecture that the generation of the mass of the first two generations
arises from a different source. If this new source is just another Higgs
doublet and if one assumes that the new doublet is neutral with respect to
the symmetry S, then one is led to a flavour structure analogous to what one
encounters in a class of the BGL-type models \cite{Branco:1996bq}, \cite%
{Botella:2009pq}, which have been extensively analysed in the literature 
\cite{Botella:2011ne}, \cite{Botella:2012ab}, \cite{Bhattacharyya:2013rya} 
\cite{Botella:2014ska}, \cite{Bhattacharyya:2014nja}, \cite{Botella:2015hoa},
 \cite{Sher:2016rhh}. If, on the other hand, the new contribution arises in
the framework of an effective field theory where the New Physics (NP)
particles have been integrated out, then assuming that this NP contribution
is of order $m_s$ and $m_c$ in the down and up sectors, one can estimate the
couplings of the Standard Higgs to $t\overline{t}$, $b\overline{b}$, $c%
\overline{c}$, $s\overline{s}$. It turns out that the couplings to $t%
\overline{t}$, $b\overline{b}$ do not differ much from those in the SM, but
the couplings to $c\overline{c}$, $s\overline{s}$ are significantly enhanced
with respect to those in the SM. \newline



\textit{Mixing in the EC limit:} We analyse quark mixing in the EC limit,
where the quark mass matrices $M_{d}$ and $M_{u}$ are rank one matrices
generated by two independent Yukawa coupling matrices $Y_{d}$, $Y_{u}$.
Therefore, $M_{d}$, $M_{u}$ can be written: 
\begin{equation}
M_{d}={U_{L}^{d}}^{\dagger }\ \mbox{diag}(0,0,m_{b})\ {U_{R}^{d}},\qquad
M_{u}={U_{L}^{u}}^{\dagger }\ \mbox{diag}(0,0,m_{t})\ {U_{R}^{u}}
\label{loo}
\end{equation}%
One does not loose generality by considering the specific ordering of $m_{b}$%
, $m_{t}$ in Eq.~(\ref{loo}), since a permutation changing these positions
can always be included in the unitary matrices $U_{L,R}^{d,u}$. The quark
mixing matrix appearing in the charged weak interactions is given by $V^{0}={%
U_{L}^{u}}^{\dagger }{U_{L}^{d}}$ and it is at this stage an arbitrary
mixing matrix. Taking into account that in the EC limit the first two
generations are massless, one can make an arbitrary redefinition of the
light quark masses through a unitary transformation of the type: 
\begin{equation}
W_{u,d}=\left[ 
\begin{array}{cc}
X_{u,d} & 0 \\ 
0 & 1
\end{array}
\right]  \label{lii}
\end{equation}
where $X_{u,d}$ are $2\times 2$ unitary matrices. Under this transformation $%
V^{0}$ transforms as $V^{0}\rightarrow V^{\prime }=W_{u}^{\dagger }\ VW_{d}$%
. One has the freedom to choose $X_{u,d}$ at will to diagonalize the $%
2\times 2$ upper left sector of $V^{\prime }$ leading to $|V_{12}^{\prime
}|=|V_{21}^{\prime }|=0$. Unitarity of $V^{\prime }$ leads then to the
constraint ${V^{\prime }}_{13}^{\ast }V_{23}^{\prime }=0$. One can then
choose, without loss of generality, $V_{13}^{\prime }=0$ and $V_{CKM}$
becomes then an orthogonal matrix, with mixing only between the second and
third generation, characterised by an angle $\alpha $, with $|V_{23}^{\prime
}|=|V_{32}^{\prime }|=|\sin \alpha |$. The important point that we wish to
emphasise is that this mixing in the EC limit of the SM, is arbitrary. The
smallness of $|V_{13}|^{2}+|V_{23}|^{2}$ in the SM, in general, cannot be
related to the smallness of the mass ratios $m_{i}^{2}/m_{3}^{2}$ where $%
i=1,2$. Therefore, in the framework of the SM the observed smallness of $%
|V_{23}|\approx 10^{-2}$, provides a hint for the presence of a flavour
symmetry. \newline


\textit{An Invariant Measure of Alignment:} Experimentally one encounters in
the quark sector $V_{CKM}\approx {1\>\!\!\!\mathrm{I}}$ which corresponds to
an alignment of the quark mass matrices in flavour space. It is useful to
have an invariant measure of the mixing defined in terms of the mass
matrices when written in an arbitrary weak basis. This can be done by
defining the following weak basis invariant \cite{Branco:2011aa}: 
\begin{equation}
A\equiv \frac{1}{2}trB^{2},\quad \mbox{with}\quad B=h_{d}-h_{u}  \label{a}
\end{equation}
where the build blocks are the two matrices: 
\begin{equation}
h_{d}=\frac{H_{d}}{tr[H_{d}]},\qquad h_{u}=\frac{H_{u}}{tr[H_{u}]}  \label{h}
\end{equation}
with the notation $H_{u,d}\equiv M_{u,d}M_{u,d}^{\dagger }$. By
construction, one has $trh_{d}=trh_{u}=1$. Given the two rank one matrices $%
M_{d,u}$, described before, corresponding to the EC limit one obtains: 
\begin{equation}
A\equiv \frac{1}{2}trB^{2}=|V_{23}|^{2}+|V_{13}|^{2}  \label{que}
\end{equation}
The result of Eq.~(\ref{que}) is exact in the EC limit. The invariant $A$
still gives a measure of the size of mixing when the first two generations
acquire mass, and in this case we have $A\approx
|V_{23}|^{2}+|V_{13}|^{2}+O(m_{s}/m_{b})^{4}$ \newline


\textit{Obtaining Small Mixing Through a Symmetry:} As stated before, mixing
in the EC limit is parametrised by an arbitrary mixing angle involving two
generations. In general, in the SM there is no reason to assume that this
mixing angle is either close to zero or maximal, in fact it can take any
value. It is possible to introduce a symmetry which leads to the vanishing
of this mixing. Without loss of generality, this angle can parametrise
mixing between the second and the third generations. Let us consider the
following symmetry S, in the context of the particle content of the SM, with
only one Higgs doublet. 
\begin{equation}
Q_{L3}^{0}\rightarrow \exp {(i\tau )}\ Q_{L3}^{0}\ ,\quad
u_{R3}^{0}\rightarrow \exp {(i2\tau )}u_{R3}^{0}\ ,\quad \phi \rightarrow
\exp {(i\tau )} \phi \ , \quad \tau \neq 0, \pi  \label{S symetry up quarks}
\end{equation}
where $Q_{Lj}^{0}$ is a  left-handed quark doublet and $\Phi$ is
the Higgs doublet. All other fermions transform trivially under S. This
symmetry leads to the following pattern of texture zeros for the Yukawa
couplings: 
\begin{equation}
Y_d = \left[%
\begin{array}{ccc}
0 & 0 & 0 \\ 
0 & 0 & 0 \\ 
\times & \times & \times%
\end{array}%
\right], \qquad Y_u = \left[%
\begin{array}{ccc}
0 & 0 & 0 \\ 
0 & 0 & 0 \\ 
0 & 0 & \times%
\end{array}%
\right]  \label{ydyu}
\end{equation}
which clearly lead to $V_{CKM}$ equal to the identity. The matrices of Eq.~(%
\ref{ydyu}) are written in the WB chosen by the symmetry. The matrix $Y_d$
can be written in the same form as $Y_{u}$ by means of a rotation of the
right-handed down quarks, which simply corresponds to a different choice of
WB.

Next we present possible ways of extending this scenario in order to
generate the masses of the first two generations of quarks, as required
experimentally, without generating large mixing and keeping the Higgs
mediated flavour changing neutral currents (HFCNC) under control. \newline


\textit{Generating the Masses of the First Two Quark Generations:} At this
stage, one has to address the question of the origin of the masses of the
first two generations. The discovery of the Higgs particle at LHC and the
study of its production and decay has shown that the vev of the Higgs field
gives the dominant contribution to the masses of the fermions of the third
generation, namely to the top and bottom quarks, as well as the $\tau $%
-lepton. It is conceivable that the masses of the quarks of the first two
generations arise from a different source, \cite{Babu:1999me}, \cite%
{Giudice:2008uua}, \cite{Goudelis:2011un}, \cite{Perez:2015aoa}, \cite%
{Altmannshofer:2015esa}, \cite{Ghosh:2015gpa}, \cite{Bauer:2015kzy}, so that
the quak mass matrices have the form: 
\begin{equation}
M=M^{(0)}+M^{(1)}  \label{m}
\end{equation}
where $M^{(0)}$ is generated by the vev of the standard Higgs $\phi $ and $%
M^{(1)}$ may arise from the vev of a second Higgs $\phi ^{\prime }$ or from
other unspecified source. In either case, the fact that there are two
different sources giving contributions to the mass of quarks of a given
charge, leads to scalar mediated flavour-changing neutral currents (FCNC).
These currents are naturally suppressed in both of the scenarios we consider
below, once the experimental values of  the $V_{CKM}$ entries 
are taken into account. \newline


\textit{Adding a Second Higgs Doublet:} The simplest possibility to generate
masses for the first two generations, is through the addition of a second
doublet $\phi ^{\prime }$ which is neutral under S. In this case, the
contribution of $\phi ^{\prime }$ to the quark mass matrix is of the form: 
\begin{equation}
M_{d}^{(1)}=\frac{v^{\prime }}{\sqrt{2}}\left[ 
\begin{array}{ccc}
\times & \times & \times \\ 
\times & \times & \times \\ 
0 & 0 & 0%
\end{array}%
\right] ;\qquad M_{u}^{(1)}=\frac{v^{\prime }}{\sqrt{2}}\left[ 
\begin{array}{ccc}
\times & \times & 0 \\ 
\times & \times & 0 \\ 
0 & 0 & 0
\end{array}
\right]  \label{m1}
\end{equation}
This structure coincides with what one encounters in a class of BGL models 
\cite{Branco:1996bq}. It has been shown that in this model, the full flavour
structure only depends on $V_{CKM}$ and thus the model obeys the Minimal
Flavour Violation \cite{Buras:2000dm}, \cite{D'Ambrosio:2002ex}, \cite%
{Botella:2009pq} principle. In this model there are FCNC but they are
naturally suppressed by small $V_{CKM}$ elements.

In this context, there are two types of BGL models: (1) top models described
by Eqs. (\ref{S symetry up quarks}),  (\ref{ydyu}) and (\ref{m1}) with FCNC only in
the down sector and (2) bottom models with the r\^{o}le of up and down quarks
interchanged. This second class of models give rise to FCNC's only
in the up sector. From low energy flavour data the scale of new physics in
top models can be quite light at a few hundred GeV \cite{Botella:2014ska}.
Bottom like models introduce new scales close to the TeV region 
\cite{Botella:2014ska}. Flavour conserving or flavour blind Higgs observables can
be accommodated in both categories because the new couplings, compared to the 
SM couplings, i.e., the coupling modifiers $\kappa_{Z}$ , 
$\kappa_{W}$, $\kappa_{t}$, $\kappa_{\tau }$, $\kappa_{b}$, $\kappa_{g}$ and 
$\kappa_{\gamma }$ may deviate from 1 at the percent level. 
These models have been extensively studied in the literature 
\cite{Bhattacharyya:2013rya} 
\cite{Botella:2014ska}, \cite{Bhattacharyya:2014nja}, \cite{Botella:2015hoa},
 \cite{Sher:2016rhh}.
In the Higgs sector the most relevant prediction specific to top models is
the decay  $h\rightarrow b\bar{s}+s\bar{b}$ with branching ratios
at most between $10^{-3}$ and $10^{-2}$ \cite{Botella:2015hoa} 
The bottom models predict the rare top decay $t\rightarrow hc$ 
with a branching ratio of at most  $10^{-3}$ \cite{Botella:2015hoa} 
In both classes of models these predictions can be correlated
with $h\rightarrow \mu \bar{\tau}+\tau \bar{\mu}$ occuring at a 
branching ratio which can reach at most $10^{-2}$


\textit{Generating light quark masses from New Physics at a high energy
scale:} Here, we consider that only one Higgs doublet is introduced in the
framework of the SM and introduce the symmetry $S$ of
Eq.~(\ref{S symetry up quarks}) which implies that only the
third generation of quarks acquire mass, with $V_{CKM}={1\>\!\!\!\mathrm{I}}$.
We shall consider that the quark masses of the first two generations arise
from New Physics contributing to the Yukawa couplings in leading order
through an effective six-order operator of the form: 
\begin{equation}
{\mathcal{L}}_{eff}=- \left( Y_{d}^{(1)}\right) _{jk}\frac{\phi ^{\dagger
}\phi }{\Lambda ^{2}}\bar{Q}_{L_{j}}^{0}d_{R_{k}}^{0}\phi - \left(
Y_{u}^{(1)}\right) _{jk}\frac{\phi ^{\dagger }\phi }{\Lambda ^{2}}\bar{Q}%
_{L_{j}}^{0}u_{R_{k}}^{0}\tilde{\phi}  \label{eff1}
\end{equation}
The Yukawa and quark mass matrices have then the form \cite{Goudelis:2011un}: 
\begin{equation}
\sqrt{2}Y_{d,u}=Y_{d,u}^{(0)}+3Y_{d,u}^{(1)}\ \left( \frac{v^{2}}{\Lambda ^{2}}
\right) \qquad M_{d,u}=v\left[ Y_{d,u}^{(0)}+Y_{d,u}^{(1)}\ \left( \frac{%
v^{2}}{\Lambda ^{2}}\right) \right]   \label{esta}
\end{equation}%
The fact that $Y_{d,u}$ are not proportional to $M_{d,u}$ leads to Higgs
mediated Flavour-Changing-Neutral -Currents (FCNC). At this stage, it is
useful to estimate the size of the new mass scale $\Lambda $. From Eq.~(\ref%
{esta}) and taking into account that $v=174$ GeV, $m_{t}=173$ GeV one
obtains $(Y_{u}^{(0)})_{tt}\approx 1$. Assuming $Y_{u}^{(1)}\approx
(Y_{u}^{(0)})_{tt}$, one obtains $\Lambda =\left[ \frac{Y_{u}^{(1)}v^{3}}{%
m_{c}}\right] ^{1/2}\approx 2$ TeV, so the new mass scale is of the order of
a few TeV. For the down quark sector, taking into account that $m_{b}\approx
4.2$ GeV, $m_{s}\approx 0.095$ GeV, one obtains $(Y_{d}^{(0)})_{bb}\approx 
\frac{m_{b}}{v}\approx 0.02$, $Y_{d}^{(1)}\approx 0.07$. Note that in the
present framework one obtains $|V_{23}|\approx O(m_{s}/m_{b})$ but one does
not provide an explanation for the smallness of $m_{b}$/$m_{t}$. We will
show that the potentially dangerous FCNC are naturally suppressed in the
present framework. The down quark mass matrix is diagonalised by : 
\begin{equation}
U_{dL}^{\dagger }\left[ Y_{d}^{(0)}+Y_{d}^{(1)}\frac{v^{2}}{\Lambda ^{2}}%
\right] U_{dR}=\frac{D_{d}}{v}  \label{u}
\end{equation}
where $D_{d}\equiv \mbox{diag}(m_{d},m_{s},m_{b})$, with an analogous
expression for the up sector. In the quark mass eigenstate basis, the Yukawa
coupling matrix becomes: 
\begin{equation}
\sqrt{2} Y_{d}^{m}= \sqrt{2}\left( U_{dL}^{\dagger }Y_{d}U_{dR} \right) =
\frac{3D_{d}}{v}-2U_{dL}^{\dagger}Y_{d}^{(0)}U_{dR}  \label{eq12}
\end{equation}
At this stage, it is useful to write $U_{dL}^{\dagger }Y_{d}^{(0)}U_{dR}$
explicitly. Taking into account that $Y_{d}^{(0)}=\mbox{diag}(0,0,\frac{m_{b}%
}{v})$, one obtains: 
\begin{equation}
\left( U_{dL}^{\dagger }Y_{d}^{(0)}U_{dR}\right) _{jk}=(U_{dL}^{\ast
})_{3j}(U_{dR})_{3k}\frac{m_{b}}{v}  \label{eq13}
\end{equation}
with an analogous expression for the up sector.
The strength of the Higgs couplings $Y_{d}^{m}$ is controlled by Eqs.~(\ref
{eq12}), (\ref{eq13}) and one has to take into account the very strict bounds on 
flavour violating scalar couplings, which can be derived from  
$K^0 -\overline{K^0}$ , $B_d -\overline{B_d}$ ,   $B_s -\overline{B_s}$,
$D^0 -\overline{D^0}$ mixings. These bounds have been recently analysed in 
\cite{Blankenburg:2012ex}. From  $B_s -\overline{B_s}$ mixing, one derives
bounds on $| (U^\ast_{dL})_{32} (U_{dR})_{33}| $ and 
$| (U^\ast_{dL})_{33} (U_{dR})_{32}| $ which taking into account that
$|(U_{dL})_{33}| \approx 1$ and $|(U_{dR})_{33}| \approx 1$, lead to
$|(U_{dL})_{32}| \leq 1.4 \times 10^{-2}$. Similarly, one derives from 
$B_d -\overline{B_d}$  mixing the bound $|(U_{dL})_{31}| \leq 3 \times 10^{-3}$.
It is remarkable that these bounds lead to   
$| (U^\ast_{dL})_{31} (U_{dR})_{32}| \simeq  | (U^\ast_{dL})_{32} (U_{dR})_{31}| 
\leq 4.4 \times 10^{-4}$ which guarantees that the Higgs contribution to   
$K^0 -\overline{K^0}$ mixing is sufficiently suppressed, to conform to
the strict experimental bound. Flavour-changing scalar couplings in the up-sector
are controlled by   $| (U^\ast_{uL})_{3i} (U_{uR})_{3j}| $. On the other hand,
$U_{uL}$ is constrained to be in a region which can generate the observed 
$V_{CKM} = (U^\dagger_{uL}U_{dL})$. Once these constraints are taken into
account, one predicts, in the present framework, the strength of flavour-changing 
decays of the top quark, namely 
\begin{equation}
 \mbox{Br} (t \rightarrow h c )
\geq 5\times 10^{-5}
\label{brbr}
\end{equation}
It is interesting to notice that in this framework the previously
analysed new flavour changing Higgs contributions all arise from
the third column of the matrices $U_{dL}$,  $U_{dR}$, $U_{uL}$ and $U_{uR}$.

So far we have only discussed the off-diagonal
Higgs couplings. In the diagonal couplings, one has to
distinguish between the couplings of the third generation (i.e. $t\bar{t}h$
and $b\bar{b}h$) and those of the two light generations. Taking into
account that $|(U_{dL}^{\ast })_{33}(U_{dR})_{33}|\approx 1$ 
and also $|(U_{uL}^{\ast })_{33}(U_{uR})_{33}|\approx 1$ it is clear
that the couplings of the third quark generation coincide with those in the
SM. On the contrary, for the first two generations, one has a significant
enhancement by a factor of $3$, leading, for example to: 
\begin{equation}
\Gamma (h\rightarrow  q\bar{q})\approx 9\Gamma ^{SM}(h \rightarrow q \bar{q})
\label{gama} \qquad q = d, s, c . 
\label{eq16}
\end{equation}
At this stage, the following comment is in order. For the down sector, the
experimental constraints from meson mixing are very strict and the
prediction of  Eq.~(\ref{eq16}) for $q= d, s$, is solid. For $c\bar{c}$
although the enhancement of   Eq.~(\ref{eq16})  holds for most of the
allowed parameter space, there are regions of allowed parameter space, 
where the enhancement is not as strong.
So far, we have only discussed the quark sector. Note that the observed
lepton flavour mixing is large and therefore there is no motivation to opt
for the symmetry S to act in the lepton sector in a way analogous to the
quark sector, since this would lead to $V_{PMNS}={1\>\!\!\!\mathrm{I}}$ in
leading order, in contrast to experiment. We shall assume that leptons are
neutral with respect to S, which leads to couplings of the Higgs particle to
leptons which coincide with those of the SM.

Taking into account that $\Gamma ^{SM}(h\rightarrow \bar{c}c)/\Gamma
^{SM}(h\rightarrow \mbox{all})\sim 3\%$, assuming Eq.~(\ref{eq16})
and that the other relevant decay
channels do not change, we get $\Gamma (h\rightarrow \mbox{all})\approx
1.23\Gamma ^{SM}(h\rightarrow \mbox{all})$. This result gives rise to a
definitive prediction for the signal strength parameters $\mu ^{f}$ \cite%
{ATLASCMS} in the decay channels $f=\gamma \gamma ,ZZ,WW,\tau \bar{\tau},b
\bar{b}$: 
\begin{equation}
\mu ^{f}=\frac{\Gamma (h\rightarrow f)\ \Gamma ^{SM}(h\rightarrow \mbox{all})
}{\Gamma (h\rightarrow \mbox{all})\ \Gamma ^{SM}(h\rightarrow f)}\approx 0.81
\label{mu}
\end{equation}
compatible with the combined ATLAS CMS analysis \cite{ATLASCMS}. Looking at
the coupling modifiers analysis $\kappa_{f}$ we have as in the SM no modification
of the couplings to the relevant channels 
\begin{equation}
\kappa_Z = \kappa_W = \kappa_t = \kappa_\tau = \kappa_b = \kappa_g = 
\kappa_\gamma = 1
\label{k}
\end{equation}
but the large enhancement in the undetected $c\bar{c}$ channel contributes
to the so-called beyond the SM branching ratio $BR_{BSM}\sim 18.8\%$ in
perfect agreement with the $34\%$ joint upper bound from ATLAS and CMS at $%
95\%$ C.L. \cite{ATLASCMS}. \\


\textit{TeV completion:} A possible TeV 
completion of the present model, can be implemented in the framework of an extension 
of the SM  where one adds three $Q = - 1/3$ vector-like quarks, $D_{\alpha}$, and three $Q = 2/3$
vector-like quarks, $U_{\beta}$, isosinglets of $SU(2)$, to the spectrum of
the SM.

We introduce the symmetry S considered in Eq.~(\ref{S symetry up quarks}),
with the standard like quarks transforming as before. The symmetry S is
spontaneously broken by the vev of the Higgs field and we also allow for a
soft-breaking term $\overline{D}_{L\beta }d_{R3}$ with a similar soft-breaking term
for the up sector. The leading
higher order operators are the ones in Eq.~(\ref{eff1}), which, in the down sector, 
are generated through the diagram of Fig. 1. 
\begin{figure}[h]
\vspace{-3.0cm}
\begin{center}
\includegraphics[width=1.0\textwidth]{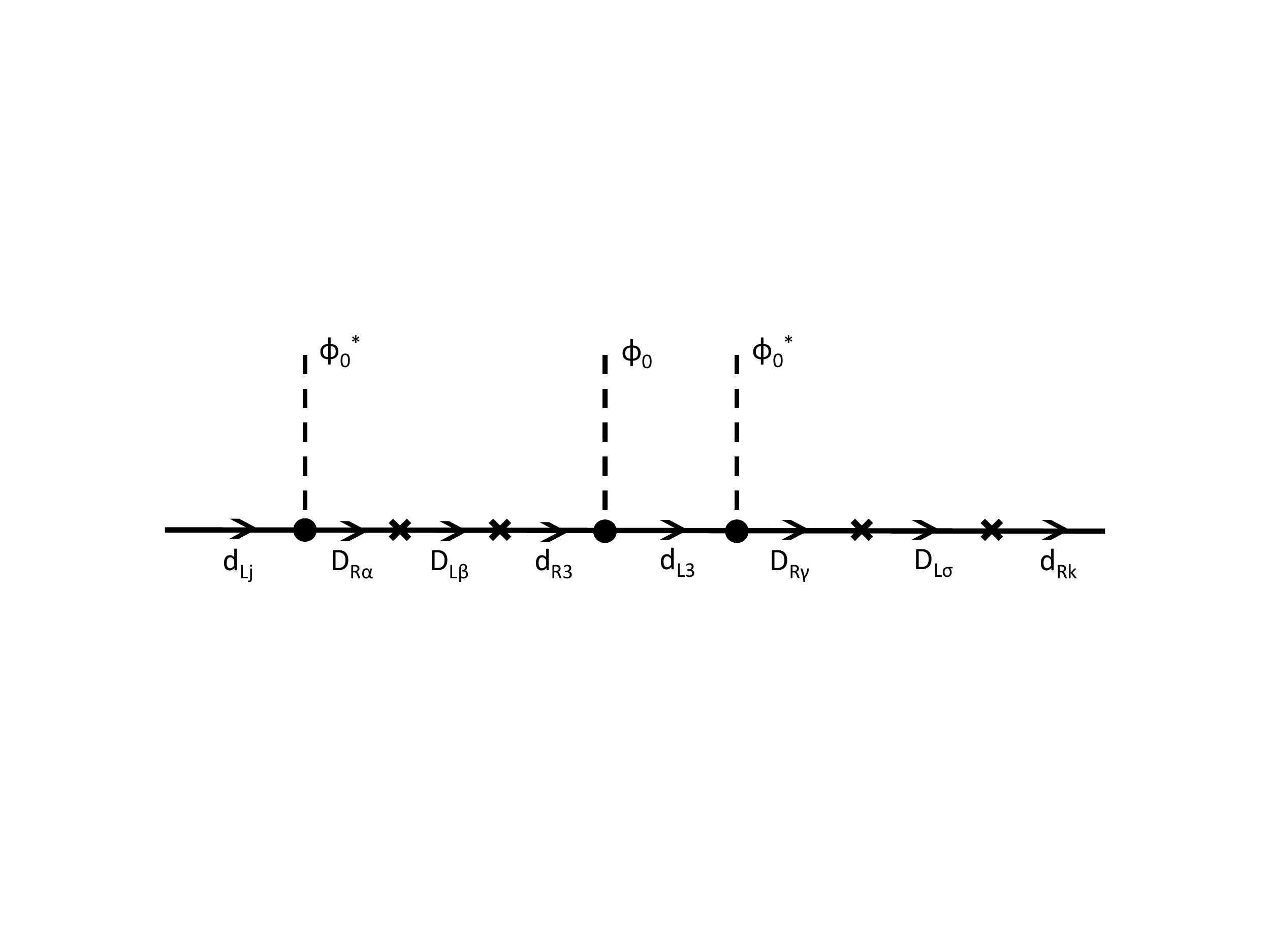}
\end{center}
\vspace{-5.0cm}
\caption{Example of generation of effective mass terms for the light down quarks after SSB,
with only one soft S breaking term involving the R component of the third generation
down quark.}
\end{figure}

It can be shown that a realistic quark mass spectrum  and a correct
pattern of quark mixing can be generated, although a full description goes
beyond the scope of this paper \cite{future}. \\

{\textit {In summary, the crucial points of our paper are:}}

- Contrary to what may be a common belief, in the SM, the natural value of
$|V_{23}|^{2}+|V_{13}|^{2}$ is of order one. In the SM, without an additional 
symmetry,  the smallness of $V_{CKM}$ mixings cannot be derived from the  observed 
strong hierarchy of quark masses.

- We point out that the fact that $|V_{23}|^{2}+|V_{13}|^{2}$ is small
may be considered as a hint of Nature suggesting the introduction of a symmetry S. 
We have given an example of such a symmetry, which leads
to $V_{CKM}={1\>\!\!\!\mathrm{I}}$ with only the third quark generation 
acquiring mass.

-We have suggested two different scenarios to generate the masses of the two lighter 
quark generations. One of them, consists of the introduction of a second Higgs 
doublet,   which is neutral under S. This framework leads to a BGL type-model 
 \cite{Branco:1996bq}, \cite{Botella:2009pq} which 
have been analysed in the literature. Another scenario consists of assuming that New 
Physics  at a high energy scale, contributes  to the light quark masses in an effective 
field theory approach. This scenario leads to the following striking predictions which 
can be tested at LHC-run 2,  as well as in other future accelerators:
The diagonal Higgs quark couplings of the 3rd generation, i.e. tth and bbh, 
essentially coincide with those of the SM. The diagonal Higgs couplings of 
the lighter  quarks are  enhanced with respect to those 
of the SM, by about a factor of three, with the most significant effect of this 
enhancement given by   Eq.~(\ref{gama}). 
In this framework one predicts Higgs mediated flavour violating top decays,
as indicated in Eq.~(\ref{brbr}) 

\section*{Acknowledgments}
The authors thank the CERN Theory Department for hospitality and partial
financial support. We thank  Luca Fiorini for interesting discussions. 
This work is partially supported by Spanish MINECO under grant
FPA2015-68318-R, and SEU-2014-0398, by Generalitat Valenciana under grant 
GVPROMETEOII 2014-049
and by Funda\c{c}\~ao para a Ci\^encia e a Tecnologia (FCT, Portugal)
through the projects CERN/FIS-NUC/0010/2015, and
CFTP-FCT Unit 777 (UID/FIS/00777/2013) which are partially funded through
POCTI (FEDER), COMPETE, QREN and EU. The authors also acknowledge the
hospitality of Universidad de Valencia, IFIC, and CFTP at IST Lisboa during
visits for scientific collaboration.

\end{document}